# Difficulties with the scars model of quantum localization

Michael J. Davis

*Chemistry Division, Argonne National Laboratory, Argonne IL 60439*
*Email: davis@tcg.anl.gov*

Numerical calculations studying bound eigenstates in chaotic regions of phase space, including those of the stadium billiard, are summarized. These calculations demonstrate that the scars of periodic orbit model is seriously flawed. An alternative picture of localization in chaotic regions is outlined.

In 1984 Heller [1] observed that some eigenstates of the stadium billiard resembled classical periodic orbits. He called this phenomenon "scars of period orbits" and developed a model to describe it. There have also been semiclassical explanations of the scarring of eigenstates, or coarse grained versions of the eigenstates [2-3], as well as mathematical studies of the behavior of the eigenstates in the semiclassical limit [4-6].

Subsequently [7-8] the model was refined and made two predictions concerning the localization of eigenstates in the stadium billiard: a numerical estimate of the amount of localization an eigenstate would have along a periodic orbit [1] and what set of periodic orbits would give rise to scars, based on their instability [7-8]. These predictions are difficult to confirm. Estimates of the degree of localization are too low by at least a factor of ten [1]. No systematic studies have confirmed the second and, in fact, several systematic studies have failed to confirm it. For example, the localization along periodic orbits in the stadium billiard can appear and then vanish as energy changes [9-10]. Scarring can also disappear as a parameter changes in directions where the periodic orbit becomes less unstable [10]. In two other chaotic systems [11-12] scars disappear altogether as energy is increased to a high enough value.

These results indicate the assumptions of the scars model need to be re-examined, and this is the purpose of this article. A re-examination is timely because of the popularity of the scars model for interpreting quantum localization in bound chaotic systems (e.g., [13]). An alternative picture of the localization is also outlined here and it is demonstrated that it can be used to understand the localization of an eigenstate in the chaotic region of an oval billiard, something the scars model is unable to do.

The two assumptions of the scars model are: 1) a wavepacket started on a periodic orbit will remain on that orbit and acquire the properties of the periodic orbit, and 2) the short time localization imposed by this process will scar the eigenstates. In order to test assumption 1, two sets of Gaussian wavepackets were propagated in the stadium billiard. One set had centers on those periodic orbits which were predicted to scar [7-8]. The other set was spread around phase space in a manner described elsewhere [14]. The centers of the wavepackets were at a "classical energy" defined by their expectation value of k, where $E = k^2/2$. The k studied here is 67.0465, which is the eigenvalue of one of the eigenstates in [1] (double diamond shape).

The calculations show that wavepackets often remain on periodic orbits for only a few bounces, and that the spreading of the wavepackets does not correspond to the schematics presented in [8]. This is demonstrated in Fig. 1, which shows the time development of a wavepacket started along the periodic orbit with a bow tie shape. The period of the orbit at the classical energy of the wavepacket is 0.100. Figure 1 shows that the wavepacket stays on the periodic orbit for short time, but after two bounces (top right) has moved off the periodic orbit and can become quite spread (bottom row). The two sets of wavepackets demonstrate that there is not a strong correlation between the recurrence properties of the wavepackets and the stability properties of the periodic orbits in contradiction to [7] (Fig. 11). In addition, these calculations demonstrate that wavepackets started on aperiodic orbits have similar recurrence properties as those started on periodic orbits [14]. These results indicate that assumption 1 is faulty.

Studies of assumption 2 have been made in [12] and [14]. There does not appear to be a strong connection between localization at short time and the eigenstates. The study of the eigenstates of the stadium billiard shows that sometimes short time localization is apparent in the eigenstates and sometimes it is not. In addition, short time localization does not depend on starting a wavepacket on a periodic orbit with short period and low instability. Wavepackets started in various regions of phase space not particularly close to these periodic orbits may have short time localization.

Figure 2 demonstrates the difficulty with assumption 2. It shows results for a coherent state [15] propagated in a two dimensional *uncoupled* harmonic oscillator, whose frequencies are not related by a small rational frequency ratio. Following [7-8], the wavepacket is propagated and its time-dependent overlap with the initial wavepacket is plotted in the top panel. The Fourier transform after short time is shown in the second panel. The structure in the spectrum leads to short time localization in the third panel. However, the eigenstates of the harmonic oscillator show no scarring. The eigenstate in the bottom panel has the largest overlap with the wavefunction above it.

Figure 2 illustrates the difficulty in connecting short time motion with the eigenstates of a system. The wavefunction in the third panel results from the wavepacket moving on a path which is essentially a torus. However, the relative ratio of frequencies is close enough to 1:1 that the wavepacket has a strong recurrence after one classical period leading to the first three panels of Fig. 2. However, this localization does not persist in the eigenstates, so there is no scarring, as demonstrated in the bottom plot of Fig. 2. The results in Fig. 2 indicate that assumption 2, like assumption 1, is faulty.

These results demonstrate that the scars model is seriously flawed. So why do many eigenstates resemble periodic orbits in chaotic regions of phase space? The apparent localization of eigenstates on periodic orbits for integrable systems, as well as chaotic ones, can be explained in terms of the structure of phase space [9,16]. Those periodic orbits which are most likely to resemble eigenstates fall into two categories [9,16-17]: 1) those which lie at the center of resonance zones (e.g., [18]) large enough to support eigenstates or are the generalization of equilibrium points, and 2) those whose manifolds form a separatrix and enclose resonance zones large enough to support eigenstates. Additional conditions for enhanced localization are discussed in [9,16-17]. Examples of the second category are presented in Fig. 3. The middle panel shows an oval billiard eigenstate [9-10] which lies in a region of phase space where the classical dynamics is chaotic. For contrast, an eigenstate of the integrable elliptical billiard is shown at the top and the bottom panel presents an eigenstate for the chaotic stadium billiard. All these eigenstates have their major density along the horizontal unstable periodic orbit.

The eigenstates of the elliptical and oval billiard can be assigned the set of Cartesian quantum numbers (28,22) by following eigenstates from lower energy. There is a separatrix in the elliptical billiard which is integrable and encloses a 1:1 resonance zone containing all classical motion which is precessional. As the separatrix gets larger with energy, progressions of eigenstates fall inside, in this case the progression (n,22). There is a non-integrable separatrix in the oval billiard which also encloses a resonance zone of all precessional motion. Once again, as the separatrix gets larger eigenstates can be captured and on their way in they resemble the separatrix. Because of tunneling [16-17], the (n,22) eigenstates resemble the periodic orbit over a range of n. A more quantitative explanation of the oval eigenstate can be made [10,16]. The area inside the separatrix at $k = 78.0133$ is 144.76 [10] and the area needed to support the (28,22) eigenstate is 144.51 ($(22 + 1.0)2\pi$, see [10,14]). This indicates that the (28,22) eigenstate lies barely inside the separatrix, and due to tunneling across the separatrix [16], the eigenstate lies on the separatrix in phase space and resembles the periodic orbit in configuration space.

There is also a non-integrable separatrix in the stadium, and the localization of the stadium eigenstate in Fig. 3 on the horizontal periodic orbit is due to tunneling across it [9-10]. However, it is more difficult to be quantitative for the stadium eigenstate in Fig. 3, although this has been done for lower energy eigenstates [10,14]. The phase space structure of the stadium is important for understanding the eigenstates [9-10], but eigenstates at higher energy move between regions, unlike less chaotic oval billiards. As these eigenstates move they resemble unstable periodic orbits, for the reasons described above. The combination of eigenstates falling into regions as they get larger, and the movement of eigenstates, should enhance the apparent localization of eigenstates along periodic orbits in the stadium.

Figure 3 shows that all the eigenstates resemble the horizontal unstable periodic orbit, but there are differences among eigenstates. For example, there is extra density in the bottom one away from the periodic orbit and there are three foci for this state compared to two for the one above it. Although the scars model has several flaws, its distraction from such important deviations may be its greatest flaw. Figure 3 shows that the localization is the common part of the stadium eigenstate, not a "scar", and results like these led to the alternative picture outlined here [10,16-17]. This alternative assumes localization and generally ignores extra density away from the localized region, so it is an approach which is the opposite of the scars model. In principle, delocalization can be included in the alternative picture by using chaotic transport models [18], which may be necessary for a better description of the stadium eigenstates, because of the movement of these eigenstates.

The goal of the work summarized here is to obtain a global description of the eigenstates of classically chaotic systems whose density of states are relevant to molecular problems (a few dozen states per degree of freedom). A global description of the eigenstates leads to the notion that some eigenstates resemble periodic orbits, but also can explain the often significant subset that does not [16-17].

It has been shown that the assumptions of the scars model are faulty, but can they be corrected? This is difficult. First, consider assumption 1. One problem is clear from Fig. 1. There the wavepacket falls off the periodic orbit and spreads wildly compared to the scars model [8]. Yet, rather than overestimating the scarring, the model is *too low* by a factor of at least 10 [1]. A second problem concerns the following. The argument made for the behavior of a wavepacket started on a periodic orbit relies on the accuracy of a particular form for this behavior which improves as $\hbar \to 0$ [1].[1] However, the scars disappear as $\hbar \to 0$ [1]. This is not necessarily a contradiction, because the limiting behavior of these two effects may be different. However, it would add complexity to the model to include this, as well as to correct the first problem.

Now consider assumption 2. Figure 2 shows there are situations where it breaks down. However, this example has strong coherence in the quantum dynamics, which some would not expect in a system with classically chaotic dynamics. But the low-lying eigenstates of the stadium are regular [21], similar to the eigenstate at the bottom of Fig. 2, so there must be considerable coherence at low energy. Although the coherence should diminish as energy is increased, estimating the extent to which it does also makes the improvement of assumption 2 complicated.

So it appears that it is difficult to correct the faulty assumptions of the scars model without losing the model's simplicity and alternative approaches should be pursued. One alternative has been presented, and it was demonstrated here and in [10,16-17] that this alternative is predictive.

This work was supported by the Office of Basic Energy Sciences, Division of Chemical Sciences, U. S. Department of Energy, under Contract No. W-31-109-ENG-38.

[1] The evidence for this lies in two papers [19-20] cited in [1].  However, one of the conditions imposed in [19-20], the differentiability of a potential function, is not met for the stadium, making the argument invalid for the stadium.

**Figure Captions**

FIG. 1. Quantum time propagation of a wavepacket in the stadium billiard started on the bow tie shaped periodic orbit. The wavepacket is propagated for a little over a period of the periodic orbit.

FIG. 2. Propagation of a coherent state of a two-dimensional oscillator leads to the autocorrelation function in the top panel, a smoothed spectrum which is the Fourier transform of the autocorrelation function in the second panel, and a short time wavefunction in the third panel. This wavefunction can be compared to the quantum eigenstate which has the largest overlap with the wavefunction (bottom panel).

FIG. 3. A series of separatrix eigenstates are plotted. The top shows an eigenstate for the elliptical billiard, the middle one for the oval billiard, and the bottom one for the stadium.